# Up-And-Coming Physical Concepts of Wireless Power Transfer

Mingzhao Song[1,2] [*], Prasad Jayathurathnage[3], Esmaeel Zanganeh[1], Mariia Krasikova[1], Pavel Smirnov[1], Pavel Belov[1], Polina Kapitanova[1], Constantin Simovski[1,3], Sergei Tretyakov[3], and Alex Krasnok[4] [*]

[1]*School of Physics and Engineering, ITMO University, 197101, Saint Petersburg, Russia*

[2]*College of Information and Communication Engineering, Harbin Engineering University, 150001 Harbin, China*

[3]*Department of Electronics and Nanoengineering, Aalto University, P.O. Box 15500, FI-00076 Aalto, Finland*

[4]*Photonics Initiative, Advanced Science Research Center, City University of New York, NY 10031, USA*

*e-mail: m.song@metalab.ifmo.ru, akrasnok@gc.cuny.edu

## Abstract

The rapid development of chargeable devices has caused a great deal of interest in efficient and stable wireless power transfer (WPT) solutions. Most conventional WPT technologies exploit outdated electromagnetic field control methods proposed in the 20th century, wherein some essential parameters are sacrificed in favour of the other ones (efficiency vs. stability), making available WPT systems far from the optimal ones. Over the last few years, the development of novel approaches to electromagnetic field manipulation has enabled many up-and-coming technologies holding great promises for advanced WPT. Examples include coherent perfect absorption, exceptional points in non-Hermitian systems, non-radiating states and anapoles, advanced artificial materials and metastructures. This work overviews the recent achievements in novel physical effects and materials for advanced WPT. We provide a consistent analysis of existing technologies, their pros and cons, and attempt to envision possible perspectives.

Wireless power transfer (WPT), i.e., the transmission of electromagnetic energy without physical connectors such as wires or waveguides, is a rapidly developing technology increasingly being introduced into modern life, motivated by the exponential growth in demand for fast and efficient wireless charging of battery-powered devices. The first WPT systems date back to the end of the XIX century and root in the foreseeing ideas of N. Tesla[1–3]. Today, WPT systems are of prime importance for many vital applications, including electric vehicles, cardiac pacemakers, and consumer electronics. Simultaneously, the amount of publications in this area has been continuously growing[4–8].

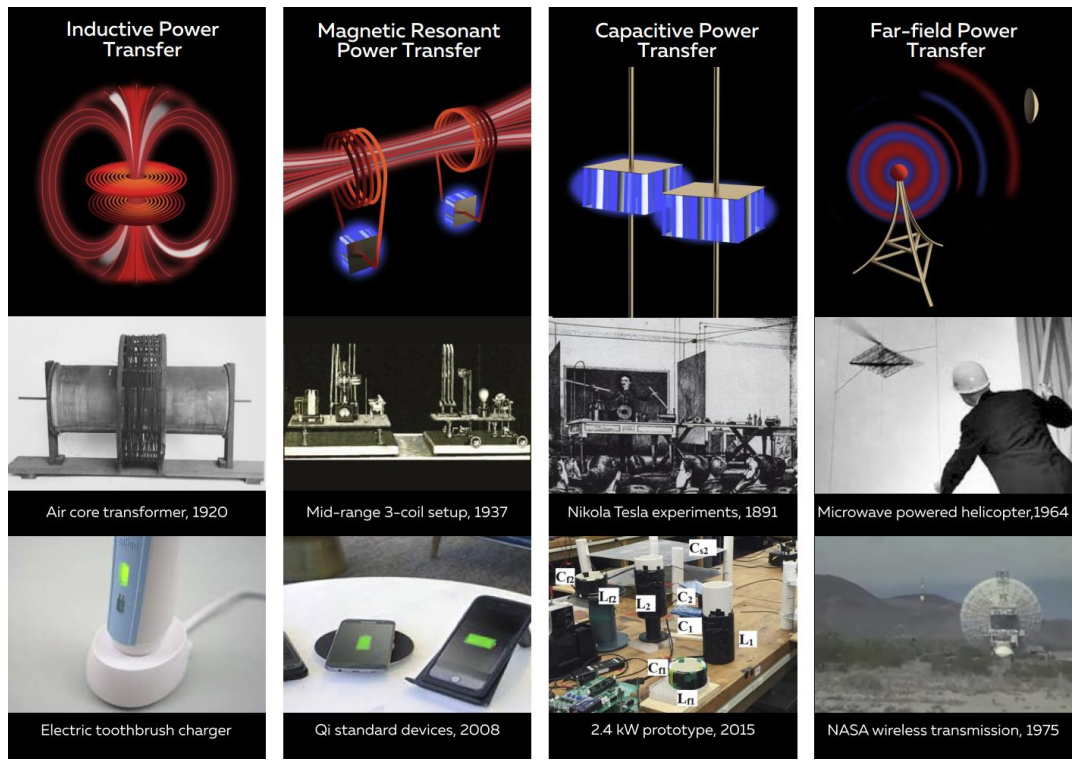

**Fig. 1 | Conventional physical mechanisms of WPT along with their classical and modern applications.** Four columns represent the basic mechanisms: inductive power transfer (IPT), magnetic resonant power transfer (MRPT), capacitive power transfer (CPT), and far-field power transfer (FPT). The first-row images illustrate the operation principles where the red colour is used for magnetic flux lines, and the blue colour corresponds to electric flux lines. The second and third rows show images of devices where the corresponding mechanisms are implemented.

The modest progress in the WPT field has been boosted since 2000, motivated by the rapid increase and qualitative development of rechargeable batteries (e.g., lithium-ion batteries) along with electronic gadgets, wireless and medical technologies, industrial and personal mobility devices. However, these conventional WPT technologies exploit mostly outdated electromagnetic field control methods proposed in the 20th century, wherein some important parameters are sacrificed in favour of the other ones. For example, high efficiency cannot be achieved together with high stability and high level of the transferred power, especially when the transfer distance is rather long. There are other restrictions (related to the direction of the power transfer, the operation frequency band, etc.), making available WPT systems far from the optimal ones.

Recently, some new methods of electromagnetic field control have been proposed. The examples include coherent perfect absorbers, bound states in continuum with arbitrarily large Q-factors, exceptional points in parity-time (PT) symmetric systems, non-radiating states, and many others[9]. On the other hand, the last decade has been accompanied by the development of new artificial materials – metamaterials and metasurfaces – that have changed researchers' approach to materials design[10–14]. The new effects and materials have led to the appearance of novel mechanisms of WPT. These mechanisms include novel effects of light-matter interaction such as PT-symmetry[15–17], exceptional points in non-Hermitian systems[18,19], coherent perfect absorption[20–22], and utilization of novel types of materials: metamaterials and metasurfaces. Efficient WPT systems via conversion of electromagnetic waves into waves of other nature (for example, acoustic waves) have also been explored[23,24]. WPT systems based on these novel mechanisms gain more and more attention, resulting in many conceptual works that may be of interest to the broader scientific and electrical engineering community. All this motivates us to review the state of the art of this actively growing area with the stress to the underlying physics and provide viable perspectives.

**Conventional WPT systems**

Let us begin by discussing the conventional WPT systems (see Fig. 1). The most well-known and, probably, most widely used WPT mechanism is *inductive power transfer* (IPT), Fig. 1(1st column). IPT utilizes two coils (transmitter and receiver) closely situated to each other[7]. An alternating current through the transmitter coil creates an oscillating magnetic field by Ampere's law. The magnetic field passes through the receiving coil, where it induces an alternating voltage by Faraday's law of induction, which creates an alternating current in the receiver.

This system and most systems discussed in what follows can usually be modelled in terms of $\hat{S}$-matrix[9], which mutually links the complex amplitudes of incoming $\mathbf{s}^+ = \{s_1^+, s_2^+, ...\}$ and outgoing waves $\mathbf{s}^- = \{s_1^-, s_2^-, ...\}$ in the form of $\mathbf{s}^- = \hat{S}\mathbf{s}^+$ (Box. 1). The amplitudes are usually normalized such that $|s_m^+|^2$ and $|s_m^-|^2$ give the power of the incoming and outgoing waves at a certain port $m$. One-port $\hat{S}$-matrix can describe the part of a WPT system along with the useful load connected to the source that delivers energy into it. In this case, delivering maximal energy from the source to the load consists in achieving the critical coupling regime[25,26], i.e., conjugated matching of the WPT system with the source impedance. A two-port network is used to describe the part of the WPT system between the source and the load: $\hat{S} = \begin{pmatrix} r_{11} & t_{12} \\ t_{21} & r_{22} \end{pmatrix}$, where $r_{ii}$ and $t_{ij}$ stand for the corresponding reflection and transmission coefficients at ports 1 and 2. In this framework, port 1 corresponds to the WPT transmitter input and port 2 is the receiver output. Thus, we can then write the expression for the *power transfer efficiency* (PTE) as a ratio of the transferred power ($|t_{12}|^2$) to the delivered amount of power ($1-|r_{11}|^2$) from the source:

$$PTE = \frac{|t_{12}|^2}{1-|r_{11}|^2}, \qquad (1)$$

assuming that all transferred power is absorbed by the useful load (critical coupling). In general, the transferred power depends on the load impedance. It is important to distinguish between PTE (so-called power gain in the microwave engineering context) and overall power transfer efficiency (the ratio of the transferred power to the power drawn from the source): PTE can reach 100% (all energy delivered to the system is absorbed in the useful load), whereas the overall power transfer efficiency never exceeds 50% when working in the critical coupling regime. Realistic WPT systems often work in unmatched conditions due to the small internal impedance of the source (e.g., switched mode power converters belong to this category) but with very high PTE. The PTE of IPT systems strongly depends on the coils' separation, relative size, and mutual orientation. If the two coils of the same size are placed close enough on the same axis, almost all the magnetic flux from the transmitting coil passes through the receiving coil, and the PTE of ~100% can be reached. In practical systems, like electric toothbrushes and shavers, the coils are of different sizes and not maximally coupled and PTE drops to 40-50%.

To overcome the distance limitation without reducing the PTE, it is possible to complement both transmitting and receiving coils by capacitances so that both these circuits experience resonance at the central frequency of a narrow operation band. This technique is called *magnetic resonant power transfer* (MRPT), Fig. 1(2nd column). Interest in this kind of WPT has been boosted after the experimental demonstration that in the near-field strong coupling regime, two resonant coils provide power transfer with 40% efficiency in the 10 MHz range, sufficient for lighting up a bulb at a distance over 2 m[27]. This study attracted strong public attention, and many scientific groups and manufacturers began to utilize it actively. Many theoretical and experimental studies of improving the PTE, matching conditions, transferred power, and the proposed system's robustness have been published so far. This method's advantages have been used to create practical WPT systems for a wide range of applications, including charging mobile and handheld wireless devices. Nevertheless, having a modest efficiency, these systems are often unstable to variations and displacements of constituting elements, limiting the charging of multiple devices.

Energy can also be transferred through high-frequency electric fields using a capacitive coupling, Fig. 1(3rd column). This method is referred to as *capacitive power transfer* (CPT)[28,29]. Here, the coupling between the receiving and the transmitting part is realized via an electric field suitable for the efficient transfer of high electric power. For example, a 2.4 kW CPT system has been designed with an air gap of 150 mm between the plates and demonstrated the PTE of 90.8%[30]. This type of WPT has been suggested for charging robots, drones, and electric vehicles. Unlike IPT and MRPT, CPT does not generate eddy current losses in conductive objects and introduces fewer fire hazards. However, CPT suffers from stronger parasitic interaction with objects located in the ambient. As a result, existing CPT solutions are targeted at lower power applications, such as coupling of power and data between integrated circuits[31] or transmitting power and data to biomedical signal instrumentation systems[32,33].

The IPT, MRPT, and CPT techniques use near-field power transfer mechanisms, where the distance between the transmitter and receiver is much smaller than the wavelength. An alternative approach is the *far-field power transfer* (FPT)[34–36], Fig. 1(4<sup>th</sup> column), where the role of the electric and magnetic fields in the energy transfer is equivalent since the energy is carried by electromagnetic waves. Conventional FPT implies power transferred by short radio waves. Microwave and mm-wave WPT systems comprise very directive transmitting and receiving antennas with a tracking system at the movable receiving part. Practically, such WPT systems offer power sufficient to support a lightweight drone up to the distances of dozens of km from the transmitting antenna. Microwave power sufficient to support all functions of a practical satellite, i.e., transferred to hundreds of km from Earth, is reported in ref.[37]. The apparent drawbacks of such an approach are the low transfer efficiency and the high density of electromagnetic energy in the space between the receiver and the transmitter. WPT carried by millimetre-range waves and microwaves is reviewed, e.g., in ref.[37]. WPT systems using light waves (high-power lasers) are also known.

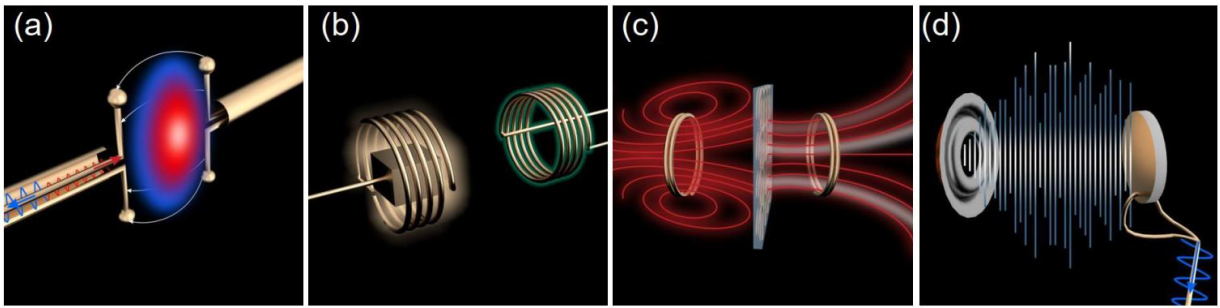

**Fig. 2 | Discussed novel power transfer mechanisms**. (a) Coherently enhanced WPT. (b) Exceptional point WPT, PT-symmetry-based WPT, and on-site power generation. (c) WPT enhanced by metasurfaces. (d) Acoustic power transfer.

**New concepts of WPT**

Fig. 2 illustrates the novel physical mechanisms of WPT that we review in what follows. Fig. 2a illustrates WPT techniques utilizing coherent perfect absorption (CPA). This approach is based on an additional excitation of the receiving antenna coherently to the field of transmitting antenna[20]. This approach uses recently developed ideas in the field of coherent perfect absorbers[21,38], and it is an example of how new effects in the field of electromagnetic field theory allow developing electromagnetic systems with better performance. Fig. 2b illustrates PT-symmetric WPT systems and systems with so-called exceptional points that exploit the physics of non-Hermitian structures. The evolution of an isolated PT-symmetric system is governed by a PT-symmetric Hamiltonian, which is symmetrical upon simultaneous reversal of space (P) and time (T)[39,40]. A PT-symmetric system consists of gainy and lossy parts that replace each other after P-symmetry operation, and then T-inversion returns it into the initial configuration. PT-symmetric Hamiltonians support a phase transition in the eigenvalue spectrum from fully real to complex ones at the so-called *exceptional points* (EPs)[9,41,42] (coalescence of eigenstates and relative eigenvalues of several modes). PT-symmetric structures are of particular interest for ellectronics[43,44] and WPT applications because they enable systems with responses robust to parameter variations[15,45,46]. More general considerations of WPT systems with balanced gain and loss lead to the concept of self-oscillating systems (on-site power generation) where the load is a part of the pulsed self-oscillating transmitter, which offer better efficiency and robust operation with variable loads[17].

Modern WPT systems possess enriched functionalities when employing metamaterials and metasurfaces[47–58]. Metasurfaces, two-dimensional thin-film structures with a subwavelength granularity, have attracted significant attention in the last few years because they offer new wave control methods in space and time, offering numerous potential applications[10,12,59–61]. Metasurfaces are used to enhance the coupling between the transmitter and the receiver of a WPT system by localizing their near fields, as illustrated in Fig. 2c. For example, inspired by the metalens concept, a self-tuning multi-Tx WPT system has been proposed, where the receiver can be randomly placed above the metasurface and get powered with high efficiency without any tracking system[62]. Metasurface itself can also be used as a transmitter in WPT systems to achieve extended functionalities owing to its capability to mold near-fields. A wisely designed metasurface can also separate an electric field from a magnetic one and provide a uniform magnetic field distribution to support charging several receivers simultaneously.

Other carriers of energy but electromagnetic fields can be used for WPT. The energy can be carried by heat flux, acoustic waves, beams of charged particles, etc. However, by definition, a WPT system transfers electric energy through an insulating ambient such as free space. Therefore, non-electromagnetic power carriers in the WPT systems can only be intermediates, i.e., direct and reverse conversion processes are implied in such types of WPT. Since these conversion processes for any practical WPT system must be energy-efficient, heat fluxes and beams of particles are not reasonable. Meanwhile, acoustic waves having very high electric conversion efficiency[63] are promising for advanced WPT techniques, Fig. 2d.

# Box 1 | S-matrix and scattering anomalies

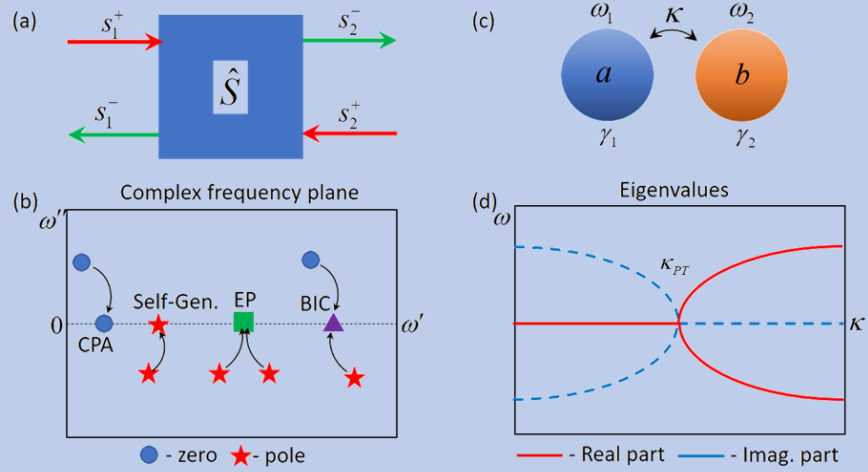

**Box 1 | Illustration of scattering anomalies.** (a) Description of a linear electromagnetic structure by a $\hat{S}$-matrix. (b) Plethora of scattering anomalies in the framework of the poles and zeros on the complex frequency plane ($\omega = \omega' + i\omega''$). CPA stands for coherent perfect absorption; Self-Gen. stands for a self-sustained generation; EP stands for exceptional point; BIC stands for the bound state in the continuum (also known as embedded eigenstate). (c) Coupled mode scenario illustrating the formation of an exceptional point and strong coupling regime. (d) Corresponding eigenvalues of the coupled-mode system as a function of the coupling strength $\kappa$.

Conventional WPT systems can usually be modelled as a two-port network, and the electric properties of components inside the network can be described by $\hat{S}$-matrix[9]. In the general case of multiple-port networks $\hat{S}$-matrix mutually links the complex amplitudes of incoming $\mathbf{s}^+ = \{s_1^+, s_2^+, ...\}$ and outgoing waves $\mathbf{s}^- = \{s_1^-, s_2^-, ...\}$ in the form of $\mathbf{s}^- = \hat{S}\mathbf{s}^+$. The amplitudes are normalized such that $|s_m^+|^2$ and $|s_m^-|^2$ denote the power of the incoming and outgoing waves at a certain port $m$. In reciprocal systems, the scattering matrix obeys the reciprocity principle, $\hat{S} = \hat{S}^T$ [64–66], for example, $S_{12} = S_{21}$, $S_{13} = S_{31}$, etc. If this relation is violated, for example, in systems with a DC magnetic field bias, time modulation, synthetic momentum, then $S_{ij} \neq S_{ji}$ for some $i$ and $j$, $i \neq j$. In a two-port network (panel **a**), the system can be described by the scattering matrix $\hat{S} = \begin{pmatrix} r_{11} & t_{12} \\ t_{21} & r_{22} \end{pmatrix}$, where $r_{ii}$ and $t_{ij}$ stand for the corresponding reflection and transmission coefficients at and between the ports 1 and 2, respectively. The $\hat{S}$-matrix can be diagonalized, $\hat{S} = \mathrm{diag}(d_1, d_2, ..., d_m)$, with $d_p(\omega)$ being the matrix eigenvalues associated with the corresponding eigenvectors $\mathbf{s}_p^+$, $\hat{S}\mathbf{s}_p^+ = d_p \mathbf{s}_p^+$. In Hermitian systems (no gain or material loss), the scattering eigenvalues $d_p$ of the $\hat{S}$-matrix are unimodular, $|d_p(\omega')|^2=1$ at any real frequency $\omega'$, which is imposed by the unitarity of the scattering matrix, $\hat{S}\hat{S}^+ = \hat{I}$ [67,68]. In non-Hermitian (lossy or gainy) systems, the scattering matrix is not unitary and eigenvalues $d_p$ can take other values $|d_p(\omega')|^2 \leqslant 1$ (lossy system) or $|d_p(\omega')|^2 \geqslant 1$ (gainy system). In the *complex frequency plane*, $\omega = \omega' + i\omega''$, scattering eigenvalues $|d_p(\omega')|$ are not bounded. The eigenvalues of the $\hat{S}$-matrix of any linear system are analytic in the complex frequency plane with a finite set of singular points, zeros and poles, see, e.g., ref.[9]. Often, the harmonic dependence on time is chosen so that the *zeros* of the $\hat{S}$-matrix of a stable system lie in the complex upper plane, and the *poles*, respectively, in the lower one. Poles of the $\hat{S}$-matrix correspond to the eigenmodes[69] of the system, whereas zeros correspond to non-scattering states (perfect absorption or trapping). It can be rigorously proven that the knowledge of these special points allows restoring all electromagnetic properties of a system under consideration[9]. Conversely, engineering these peculiarities in the complex frequency plane allows tailoring structures with unusual scattering effects (panel **b**).

In lossless (Hermitian) systems, zeros and poles are complex conjugated, i.e., symmetric to the axis of real frequencies. Adding losses to a Hermitian system and thereby turning it into a non-Hermitian system leads to the loss of this relation between zeros and poles, and one of the zeros can reach the real axis, turning the system into a *perfect absorber* (PA)[70]. If the system has two or more scattering channels, this can lead to *coherent perfect absorption* (CPA)[9,71,72]. In the linear scenario, CPA and lasing can be considered as time-reversal ($\hat{T}$) versions of one another. CPA enables a device that can completely absorb coherent input light due to the destructive interference between the transmission and reflection outside a cavity (or a resonator) and constructively interfere inside the cavity. In this case, the overall absorption and transmission can be dynamically controlled with careful tuning of the phase difference between the incident waves[9,21]. Exploiting coherence to control optical phenomena offers far richer opportunities beyond enhancing absorption. For example, instead of adding material loss to a system to achieve CPA, one can play around with radiation losses, where the coherent excitation leads to manipulation of radiation[73].

Another effect in scattering theory is a self-sustained generation, which is realized when an amplifying medium is added, which leads to the shift of a pole to the real axis. In the general theory of light scattering, this situation corresponds to the regime of a laser[74]. This regime of loss compensation can also be achieved in the PT-symmetric regime, very promising for WPT applications[17].

If the zero and the pole of a Hermitian system converge on the real axis, this state corresponds to the so-called *bound state in the continuum* (BIC)[75]. The pole near the real axis gives an unboundedly large Q-factor, which is possible due to the pole's cancellation by the zero. Recently, BICs have been found and studied in many Hermitian structures, including photonic crystals, metamaterials, and metasurfaces[75–78]. It was shown that the realization of BICs is interesting for a wide range of practical applications, especially sensing and low-threshold lasers. However, combining the zero and pole of the $\hat{S}$-matrix at one point on the real frequency axis requires an absence of losses in the system, which contradicts to the fact that all systems are lossy in one way or another. BICs being *nonradiative states* are attractive for WPT as they may enable radiation-free systems[78].

If two poles coalesce at one point, it corresponds to an exceptional point (EP). The most interesting case for applications is the real EP, possible in systems comprising gain in the PT-symmetric regime. Perhaps, the simplest example of a PT-symmetric structure is a set of two coupled resonators, which are identical in any sense except for the presence of gain in one element (a) while the other element (b) exhibits the same amount of loss (panel **c**). The equation governing these systems reads $\frac{d}{dt}\begin{pmatrix}a\\b\end{pmatrix}=-i\hat{H}\begin{pmatrix}a\\b\end{pmatrix}$, where $a$ and $b$ are the amplitudes of oscillations in two resonators, and the Hamiltonian reads[15,79–82]:

$$\hat{H}=\begin{pmatrix}\omega_{0,1}-i\gamma_1 & \kappa\\ \kappa^* & \omega_{0,2}-i\gamma_2\end{pmatrix}, \tag{2}$$

where $\omega_{0,1}$ and $\omega_{0,2}$ are the eigenfrequencies of the resonators in the absence of coupling, $\gamma_i$ are the decay/gain factors of resonator $i$, and $\kappa$ denotes the coupling strength. Solving the eigenstate problem gives the following eigenvalues ($\omega_\pm$):

$$\omega_\pm=\omega_{0,\text{ave}}-i\gamma_{\text{ave}}\pm\sqrt{|\kappa|^2+(\omega_{0,\text{dif}}+i\gamma_{\text{dif}})^2}, \tag{3}$$

where $\omega_{0,\text{ave}}=(\omega_{0,1}+\omega_{0,2})/2$, $\omega_{0,\text{dif}}=(\omega_{0,1}-\omega_{0,2})/2$, $\gamma_{\text{ave}}=(\gamma_1+\gamma_2)/2$, $\gamma_{\text{dif}}=(\gamma_1-\gamma_2)/2$. In the case when $\omega_{0,1}=\omega_{0,2}\equiv\omega_0$ and $\gamma_2=-\gamma_1\equiv\gamma$, the eigenvalues of the two modes become $\omega_\pm=\omega_0\pm\sqrt{|\kappa|^2-\gamma^2}$. The analysis of (3) shows that if the coupling is weaker than a certain critical value ($\kappa<\kappa_{\text{PT}}=\gamma$), the system possesses two modes, one lossy and one amplifying (panel **d**). When excited, the lossy mode decays exponentially in time, whereas the gainy one exhibits exponential growth. However, if the coupling strength is large enough ($\kappa>\kappa_{\text{PT}}$), the system resides in the strong coupling regime when the coherent energy exchange between the resonators compensates for the loss decay, stabilizing the system at the real frequency axis. The critical point $\kappa=\kappa_{\text{PT}}$ gives rise to an EP. Exactly at the transition threshold, $\gamma_2=-\gamma_1$, two modes coalesce to the single eigenstate: $(1,i)^{\text{T}}/\sqrt{2}$ with the eigenfrequency $\omega_0$, featuring non-Hermitian degeneracy. PT-symmetric systems and EPs provide reliable WPT with the mutual distance between the resonators in a relatively wide range[15,83].

**Scattering anomalies for WPT**

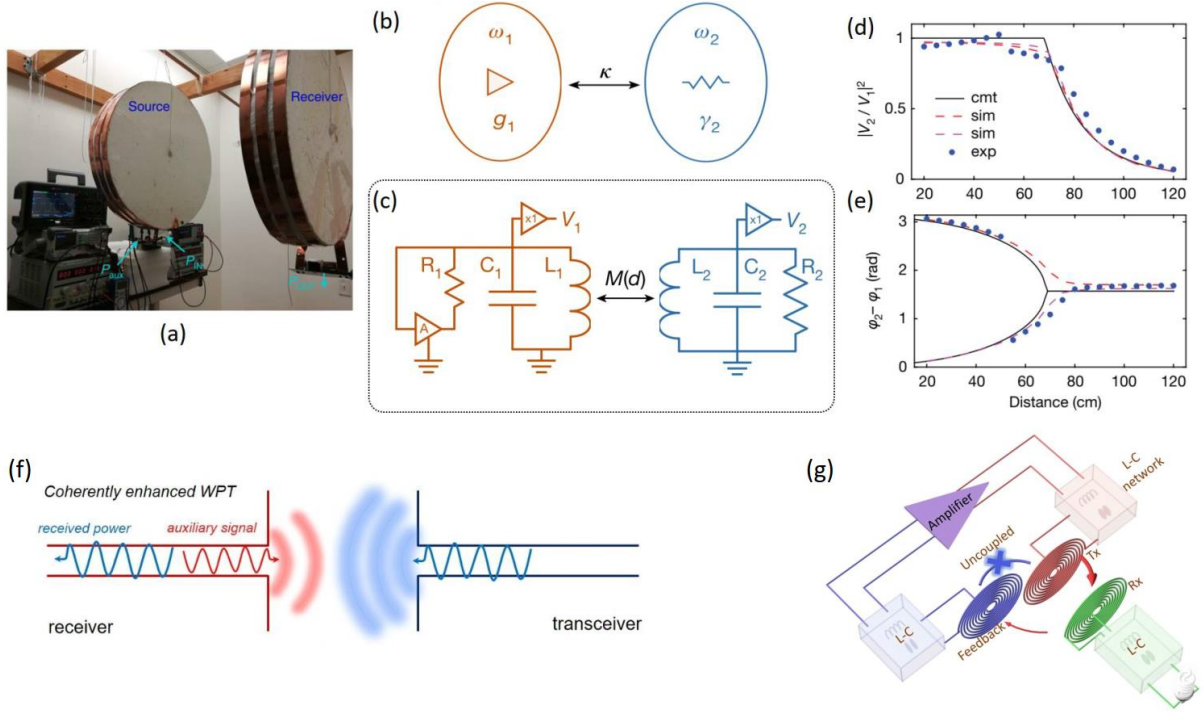

**Fig. 3 | Scattering anomalies for WPT**. (a) PT-symmetric WPT system consisting of two magnetic resonant antennas (coils). (b) Coupled-mode theory model of a pair of coupled gain-loss resonators. Triangle indicates gain; resistor symbol indicates a loss. (c) Circuit theory model for PT-symmetric wireless power transfer, showing inductor (L), capacitor (C), resistor (R) and amplifier (A); the triangle marked '× 1' is a voltage buffer. (d, e) Coupled-mode theory (cmt), circuit simulation (sim) and experimental (exp) results for voltage ratio (d) and phase (e). The two sets of simulations (red and magenta) use slightly ($\pm 5\%$) different $C_2$ and $R_1$ values. These deviations not only show the system robustness but also make the eigenfrequency bifurcation arising at $D \approx 65$ cm (PT-symmetric regime) better visible. (f) Coherently enhanced WPT system consisting in the excitation of the receiving antenna by an auxiliary coherent signal. (g) Concept of on-site wireless power generation using inductively coupled wireless links. Figure adapted from: **a-e**, ref.[15].

A major challenge for practical WPT systems is achieving stable power transfer with high and robust transfer efficiency upon variations of coupling and load conditions. Recently, this issue has been addressed with systems supporting PT-symmetry and EP (Box. 1). Real-valued EPs are possible in systems with gain and require strong coupling of modes. Let us consider a scenario depicted in Figs. 3a,b. Two resonators (in WPT terminology: transceiver/source and receiver, respectively) support (uncoupled) eigenmodes with frequencies $\omega_1$ and $\omega_2$, with decay rates $\gamma_1$ and $\gamma_2$. The modes are coupled with the coupling strength $\kappa$. The source eigenmode (in the absence of the receiver) is supposed to be gainy: $-\gamma_1 = g_1$, where $g_1$ is the gain parameter, which is assumed to be nonlinear and saturable. The Hamiltonian of this scenario reads $\hat{H} = \begin{pmatrix} \omega_1 + ig_1 & \kappa \\ \kappa & \omega_2 - i\gamma_2 \end{pmatrix}$. From now on, the real frequencies of the modes are assumed to be equal, $\omega_1 = \omega_2$. The linear analysis shows that if the coupling is lower than a certain critical value ($\kappa < \kappa_{PT}$), the system has two modes (more precisely, quasi-modes): lossy and gainy ones, appearing at the same real frequency (PT-symmetry-broken phase). However, if the coupling strength is large enough ($\kappa > \kappa_{PT}$), the system is in the strong coupling regime (PT-symmetric phase), see Box. 1. As a result, the modes share the same imaginary frequency (vanishing in the balanced PT-symmetric regime) and amplitude but acquire altered real ones because of coupling. The real EP emerges exactly at the boundary between these two regimes ($\kappa = \kappa_{PT}$). The essential point here is that the nonlinear gain parameter ($g_1$) of the source is saturable, which allows to overpump the system to achieve the real-valued EP for a relatively large distance between the source and receiver. In this regime, the amplitudes of both modes are equal, giving rise to the voltage ratio close to 100%. In ref.[15], this has been realized with a high-speed operational amplifier configured as a non-inverting amplifier (triangle A in Fig. 3c). Now, if the distance (d) between the source and receiver changes within [0,d($\kappa_{PT}$)], the PTE stays close to 100%.

The results of examining the system are shown in Figs. 3d,e. The system demonstrates the transition from the PT-symmetric phase to the PT-symmetry-broken phase at a distance between the resonators of 70 cm. In the PT-symmetry phase, the system exhibits high power transfer efficiency as both resonators form a hybrid mode and share the same voltage drop ($|V_2/V_1|^2 = 1$). Remarkably, the variation of the distance from 0 to 70 cm does not cause any significant change in the WPT efficiency. In practice, this WPT system is robust to sufficiently small mutual displacements and turns between the receiving and transmitting antennas[15].

The PTE differs from the overall power transfer, i.e., the ratio of the amount of energy delivered to the receiver to the total power drawn from the source, which never reaches 100% due to various dissipations. Moreover, this nonlinear PT-symmetry approach requires the system to be strongly pumped, leading to additional strong dissipations in the amplifier (A). In ref.[15], the amplifier was not optimized, which resulted in relatively low system efficiency of ~10%. In the subsequent work[83], the authors have used a power-efficient switch-mode amplifier with current-sensing feedback in the PT-symmetric circuit. The reported nonlinear PT-symmetric radiofrequency circuit can transfer around 10 W of power to a moving device with a nearly constant PTE of 92% over distances from 0 to 65 cm.

In ref.[16], this approach has been applied for drone-in-flight wireless charging. The reported experimental results have shown that when the flying drone hovers in a confined three-dimensional volume of space above the WPT system, the stable output power is maintained with PTE of 93.6%. In ref.[84], a frequency-locked WPT system based on higher-order PT-symmetry was proposed, where a near-unity PTE has been experimentally verified in presence of distance variation, misalignment between coils and impedance fluctuations in electric grids. PT-symmetric electronic systems can be also used for sensing. In ref.[85,86], a coherent-perfect-absorber-laser-locked PT-symmetric electronic circuitry was used to build RF sensors that are capable of detecting subtle impedance changes with high sensitivity.

In general, such PT-symmetric WPT systems can be implemented by using a self-oscillating amplifier as the power source where the eigenfrequencies are naturally tracked[87]. Alternatively, an external pulsed driven switched-mode power amplifier can also realize a similar PT-symmetric behavior[88]. In both cases, the generator acts as a negative resistor (gainy system). On the receiver-side a simple passive rectifier or a switched-mode power converter can be connected to the load or battery, which acts as a positive resistance (lossy system). We note that there are also earlier designs of WPT systems with self-oscillating switched-mode power supplies[89] that perform identically to the PT-symmetric WPT systems.

The next approach we discuss is the coherently enhanced wireless power transfer. This approach originates from the CPA effect (see Box. 1) when a system with a certain amount of loss is excited by the wave configuration corresponding to $\hat{S}$-matrix zero[21]. It was shown that the same principle can be employed to improve the PTE[20]. Namely, coherent excitation of the *receiving* antenna with an auxiliary wave, tuned in sync with the incident wave incoming from the transmitter, allows significant boosting of PTE and enhances the robustness of the system, Fig. 3f. The auxiliary wave enables destructive interference with the reflected wave, compensating for possible mismatch in the receiving antenna.

To analyse this scenario, we assume that the receiving antenna has a single mode with the real eigenfrequency $\omega_0$ and amplitude $a$. The antenna is coupled to the output waveguide (loaded by the useful resistance) and free space (where the incident signal comes from) with the coupling constants united into a two-component vector $|\kappa\rangle = \{\kappa_w, \kappa_f\}$. The antenna excitations can decay to both channels with the total decay rate $\gamma = 1/\tau_w + 1/\tau_f$. The results of the coupled-mode theory analysis of this approach yield the extracted energy[20]

$$P = \left| s_w^+ + \kappa_w \frac{\kappa_w s_w^+ + \kappa_f s_f}{i(\omega - \omega_0) + \gamma} \right|^2, \quad (4)$$

where $s_f$ and $s_w^+$ are the amplitudes of the incident waves generated by the transmitting antenna and the auxiliary wave generated in the receiver. As follows from (4), the initially excited mode decays in time as $a \sim e^{i\omega_0 t} e^{-t/\tau}$, i.e., exponentially with the complex frequency, $\omega = \omega' + i\omega''$, where $\omega' = \omega_0$ and $\omega'' = -1/\tau$. If $s_f = 0$ (no input from the transmitter), this expression gives the reflection parameter $S_{22}$ of the receiving antenna. The analysis shows that there are parameter (auxiliary wave's amplitude and phase) areas where the net extracted energy exceeds that for absent auxiliary coherent excitation. In these regimes, the antenna receives more power from the transmitter than it does without coherent excitation even after subtracting the control wave energy generated in the receiver. In a practical WPT device, the amplitude and phase of the additional coherent signal can be controlled in real-time to adjust the antenna to the changes in the environment, changes in the load, and distance from the transmitter.

The physical principle of this approach is similar to conventional impedance matching. Indeed, a usual matching element inserted between the receiving antenna and the load also sends the antenna a coherent wave with the field emitted by the transmitting antenna, leading to increased received power. The difference with the solution proposed in ref.[20] is that in conventional devices, this coherent wave is created by reflection (with the proper phase shift) of a part of the received wave, while in paper[20], it is proposed to generate it by a dedicated phase-locked oscillator which can be made tunable.

The next approach suggested recently is the so-called *on-site generation*, which is intrinsically related to the self-sustained generation regime when one pole of an active system reaches the real frequency (see Box. 1). The importance of this approach is as follows. In any WPT system, the source of power is either a DC or 50/60 Hz source. The load (usually) requires DC voltage. Thus, any WPT system must contain a generator (a self-oscillating system) that converts DC or 50 Hz into some reasonably high-frequency oscillations (either harmonic or pulsed). The conventional WPT devices realize the following conversion: generator → free-space link → receiver → rectification. In a conventional chain system, oscillations exist even if there is no receiver because the closed positive feedback loop closes inside the transmitter. In contrast to this conventional approach, in the case of on-site generation[17], the feedback closes through the receiving load, and if there is no load, oscillations are disabled, Fig. 3g. This regime may be very advantageous for MRPT and CPT because instead of transferring the energy of an alternating current through the gap between the coils or plates of two capacitors, it offers generation of the alternating signal directly in the useful load[17]. An on-site power generation paradigm, in which the wireless link and the load together form the feedback loop of a self-oscillating circuit, has been proposed to achieve robust WPT operation[17,89,90] and robust wave tunneling and information transfer[91]. With such a united self-oscillating configuration, the system variations, i.e., the load and/or receiver position, changes, modify the feedback loop, and the system can track the optimal WPT condition.

**Metamaterials and metasurfaces for WPT**

Metamaterials, artificial structures composed of electrically small unit cells (meta-atoms), allow sculpturing materials with unusual properties on demand[92], Fig.4a. A metamaterial is composed of meta-atoms as a natural material is composed of atoms, Fig.4b, but in a controllable fashion. A variety of metamaterials to improve the WPT efficiency have been recently proposed[55–57,93–102]. To achieve a unique electromagnetic response, different meta-atoms made of various materials and structures have been proposed[7,11,13,14,103,104], among which metallic structures (split-ring resonators) and dielectric structures (high-Q high-index dielectric resonators) are widely adopted due to their superb ability to achieve strong magnetic response, Fig. 4d. For example, in Ref.[104], a WPT system based on two dielectric resonators operating at the magnetic quadrupole mode was proved highly efficient than operating at lower-order modes.

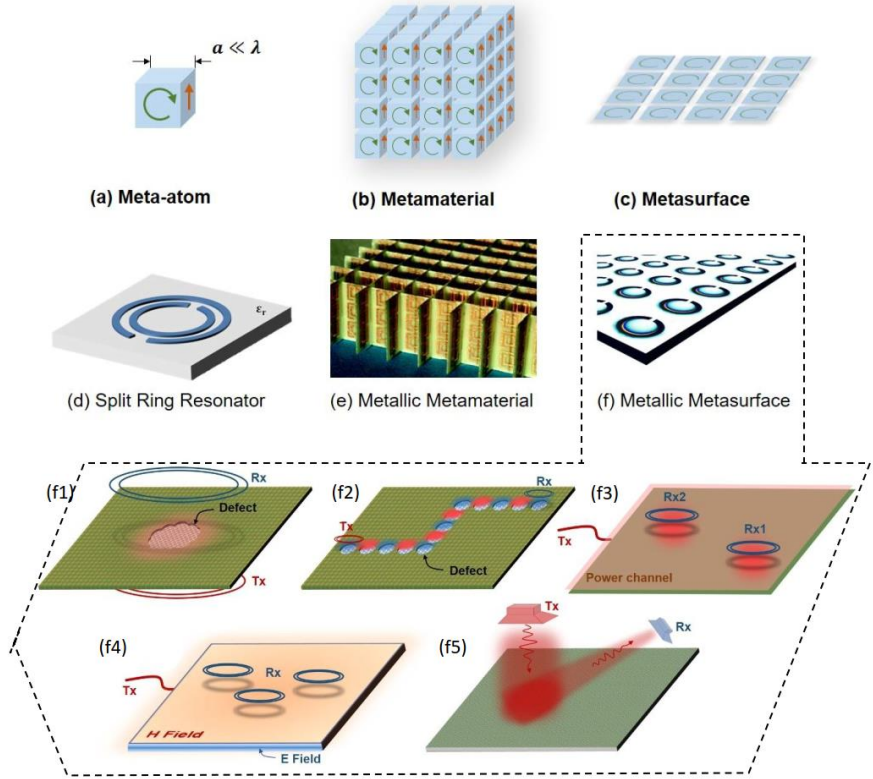

**Fig. 4 | Meta-structures in different dimensions**. Schematic figures of (a) meta-atom, (b) metamaterial and (c) metasurface. (f1-f5) Near and far-field control for WPT implemented by different metasurfaces. (f1) Cavity mode excited in the defects in the metasurface for energy concentration. (f2) Patterned defects in the metasurface to guide surface waves. (f3) Self-tuning power channels for multiple receivers. (f4) Near field modification for making the magnetic fields uniform and shielding the electric fields. (f5) Wavefront control by metasurface to redirect and focus the far-field power flow.

Compared to metamaterials that are bulky and cumbersome in many cases, their two-dimensional counterparts, *metasurfaces*, are considered good alternatives, Fig. 4c. Metasurfaces have recently attracted a great deal of interest due to their potential to influence electromagnetic fields profoundly[11]. A variety of functionalities are demonstrated for a broad frequency band ranging from the microwave region to the visible region. Recently, metasurfaces have been intensively studied for WPT applications based on their superb ability to manipulate near field and far-field. For example, in ref.[48], the *smart table concept* is proposed, where the WPT system operates on a metasurface formed by a wire array supporting surface waves. This surface wave can propagate along the wires due to the canalization effect. Therefore, the field leakage to other directions can be well confined, and the coupling between the transmitter and a distant receiver can be significantly enhanced.

A cavity formed in a metasurface has been utilized to localize the fields into a subwavelength region, Fig. 4f1. Here, the transmitter (Tx coil) and the receiver (Rx coil) are placed at both sides of the defective metasurface that is inductively coupled to the Tx and Rx resonators, mainly for the impedance step-up[58]. For a certain size ratio between Tx and Rx coils, there exists an optimal configuration of the cavity size[58]. The location, shape and intensity of the hot-spot generated around the cavity can be actively and dynamically controlled by using tunable components such as varactors[54].

The defect-based approach can be generalized to waveguide channels where the propagation control is achieved using reconfigurable defective cavities formed in the metasurface, Fig. 4f2. The physical mechanism for creating the cavity is described using Fano interference, in which the resonant frequency of the cavity falls into the bandgap of the metasurface. In ref.[105], a non-uniform metasurface is proposed to control the wave propagation using the hybridization bandgap that provides an effective suppression of the surface wave. The tunable magnetic metasurface enables localization of magnetic field in deep sub-wavelength volumes. Consequently, the transmission is significantly enhanced compared to a uniform metasurface, paving the way for directional wireless power transfer on a surface.

The potential of magnetic resonance coupling for energy transfer to multiple receivers was shown in Ref.[106], where the power was delivered simultaneously from a large Tx-coil to several small Rx-coils. PTE in these systems highly depends on the coupling coefficient between coils. As shown in Ref.[107], such systems can be optimized for a certain number and position of Rx-coils to reduce this effect.

In multi-Rx systems with magnetically coupled resonators, the concept of selective energy transfer and the simultaneous powering of diverse-standard devices was applied. It consists of multi-frequency WPT, which was carefully elaborated and implemented[108–112]. This concept can also be implemented for metasurfaces. In ref.[113], a multi-mode metamaterial-inspired resonator formed as an array of sub-wavelength parallel conductors was introduced. It was shown that the structure has several eigenmodes with different magnetic field configurations, and the efficiency of power transfer can reach 88%. This enables power devices to operate at different frequencies with a single transmitter.

A self-tuning multi-Tx WPT system capable of automatic optimal power channeling through nearest Rx and Tx coils without any sensing or control circuits has been proposed in refs.[62,114], Fig.4f3, which was inspired by the known idea of so-called superlens based on two parallel arrays of small resonant objects[115,116]. This WPT system consists of a transmitter layer and a metasurface layer, where the transmitters close the receiver are excited strongly, generating self-tuning power channels[62,114]. In this approach, the currents in individual channels are self-tuning based on the receiver's position, and the total efficiency remains almost constant throughout the complete receiver range.

Metasurfaces can also be designed so that the electric and magnetic near fields are well separated. In ref.[53], a two-layer wire array embedded in a high-permittivity background material is used as a WPT Tx resonator, as schematically shown in Fig. 4f4. It is demonstrated that the specific absorption rate (SAR) can be significantly reduced compared with the conventional spiral coil resonator. Thereby the maximal allowable power can be increased, staying within the limits of safety regulations.

The efficiency of FPT relies on highly directive transmitting and receiving antennas. Conventionally, high directivity can be achieved in large-scale phased-array antennas, which, however, are not ideal solutions for WPT due to their complicated implementations and large device dimension[117]. One of the metasurface's essential properties is its superb capability to interact with the impinging wave efficiently and form desired scattering/emission pattern, Fig. 4f5. For instance, in ref.[118], a metasurface with an area of 1.5 m$^2$ was designed consisting of over 4000 reflective patch resonators operating at 5.8 GHz. The reflective phase response of each element can be carefully tuned by changing their geometries. After a proper modification of this metasurface, the required phase distribution can be obtained. The impinging wave generated by a feeding antenna is well manipulated and focused at 6.5 m away from the metasurface in the desired direction enabling transfer efficiency is as high as 40%.

**Acoustic WPT**

Acoustic power transfer (APT) is one of the emerging branches of WPT. Here, electromagnetic energy is converted to mechanical vibrations that propagate through some media and then converted back to the receiver's electromagnetic energy[23,24], Fig. 5a. Fig. 5b illustrates a piezoelectric transducer (PZT), which is widely used in APT. PZT can be flexible and even biodegradable[119–121]. The energy conversion efficiency in PZT usually lies in the range 70 – 90%[122], determined by the electromechanical coupling coefficient. The most significant disadvantage of PZTs is the necessity of a matching layer needed to minimize wave reflection at the boundary between the transducer surface and media of propagation.

Advantages in micromachining techniques enabled new types of transducers – piezoelectric (PMUT) and capacitive (CMUT) micromachined ultrasound transducers[123–125], Figs. 5c,d. Typically, micromachined transducers have larger bandwidth and produce higher acoustic pressure in comparison to PZTs. The electromechanical coupling coefficient of CMUT and PMUT may reach 70% and 45% correspondingly[126,127]. In PMUTs, the membrane is a piezoelectric layer connected to electrodes. The incident wave causes oscillations and, therefore, the membrane's strain, which results in electricity generation. The reverse process results in wave generation. CMUT also includes a membrane, but in this case, one of the electrodes is placed on a substrate so that the capacitance of the system changes with oscillations. CMUT requires a DC bias or some permanent charge, which may be undesired for APT applications, especially for medical implants. To overcome this limitation, the concept of capacitive parametric ultrasonic transducer (CPUT) was introduced[128,129]. Preliminary models show that the efficiency of CPUT can reach 90%, making this type of transducers rather promising.

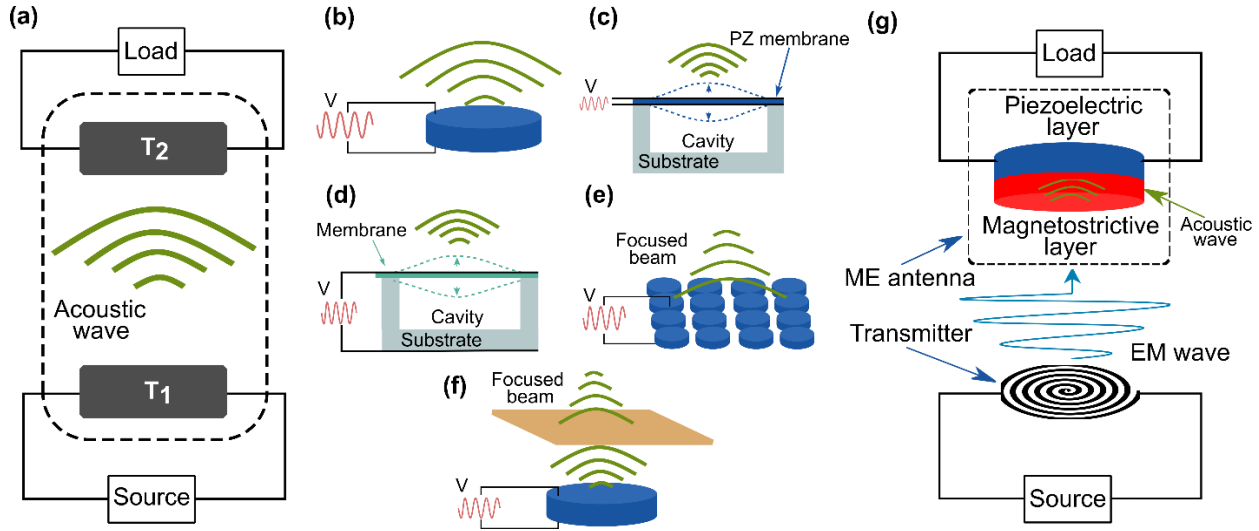

**Fig. 5 | Acoustic power transfer**. (a) General scheme of APT. (b-f) Various configurations of APT transducers: (b) Piezoelectric, (c) Piezoelectric micromachined ultrasound transducers, (d) Capacitive micromachined ultrasound transducers, (e) Phased array, (f) Acoustic hologram. (g) General scheme of WPT based on a magnetoelectric antenna (ME).

Recently, the idea of using a phased array transmitter to focus acoustic power at the receiver location and increase the efficiency of APT was suggested in ref.[130], Fig. 5e. A 37-element phased-array (each element is connected to a phase shifter) with a total diameter of 7 cm is used as a transmitter. The distance between the transmitter and receiver is about 5 times the receiver diameter. A factor 2.6 increase in the overall efficiency compared to a single transducer with the same distance was achieved at 40 kHz.

Another new approach to improving efficiency by focusing on transferred power is using acoustic holograms between the transmitter and receiver, Fig. 5f. In ref.[131], the authors proposed a 3D-printed acoustic hologram to create a multifocal acoustic pattern in a target plane where receivers are located at specific positions. The reported power transferred to the target receiver is ~2 times the power transferred to the unwanted receiver.

For decades WPT technologies have been considered for powering implantable devices. Electromagnetic devices are bulky because of their large antennas. On the other hand, acoustic-based implantable devices have low data rates and significant losses, especially in the skull. Recently, magnetoelectric (ME) antennas[132–135] based on composites of multiferroic structures have been proposed, Fig. 5g. In a receiver, when the magnetostrictive layer of the ME antenna senses the magnetic component of the electromagnetic wave, it vibrates and produces strain (acoustic wave). Because the piezoelectric layer is mechanically connected to the magnetostrictive layer, it transfers to the piezoelectric layer. These systems provide a good alternative for conventional receivers in implantable electronic devices. For instance, in ref.[136], two ultra-compact ME antennas, which are one to two orders of magnitude smaller than state-of-the-art compact antennas, have been fabricated and tested. Realizing at 60 MHz and 2.525 GHz, the ME antennas acoustic transmitting and receiving mechanism have been successfully characterized in nanoplate resonators and thin-film bulk acoustic wave resonators. Comparison studies reveal that ultrasonic systems can be more effective than electromagnetic ones, especially for small deep-implanted devices[137,138].

**Outlook**

Here, we foresee several possible pathways in the actively developing area of WPT. First, the utilization of nonradiative states, like *anapole* and *BICs*[139] (see Box. 1), could be greatly beneficial for WPT applications as they allow systems with no parasitic radiation energy loss. Anapole state is a nonscattering state of the electromagnetic field that manifests itself as a pronounced deep in scattering efficiency. Despite its non-scattering nature, the inner electromagnetic fields of an anapole state can be strongly enhanced, giving rise to boosted light-matter interaction effects[140]. Thus, the coupling of two resonators set at the anapole state can boost WPT efficiency because of vanishing radiation loss. However, it is worth noting that an anapole state is the interference of at least two multipole moments[141], and hence an alternation of the moments caused by the presence of the second resonator/antenna can distort the anapole, making the system sensitive to the change of parameters.

BIC is an eigenmode of a resonator that does not radiate because of either symmetry or destructive interference with another, not orthogonal, mode enabling scattering resonances with large Q-factors, bounded only by dissipation losses (Box. 1). BIC-supporting structures provide a unique platform for tailoring an electromagnetic environment and the radiation losses, which is of paramount importance for high-frequency WPT. Also, recent studies show that BICs are topologically protected, i.e., variation of the system parameters does not eliminate the BIC state[142]. The robustness of these BICs was explained by their topological nature[143,144] rooted in the fact that these features obey the conservation of the topological charge[145,146]. It was also shown that the merging of BIC charges enables even more confined resonances in realistic systems[147] and unidirectional guided modes within the continuum[148]. Such topological protection makes BIC-supporting structures interesting for stable WPT systems.

Next, systems with explicit time modulation attract significant attention nowadays as they allow overcoming many limitations and bounds, including Lorentz reciprocity[149,150], passive PT-symmetry[151], Chu limit[152], and nonmagnetic topological phases of matter[153,154]. For example, beating the Chu limit with explicit time modulation could enable WPT systems with ultrasmall but sufficiently wideband transceivers and receivers. The adiabatic time modulation can lead to geometric phase effects that can be used to break the time-reversal symmetry and create various topological effects[155–157]. All this make systems with time-modulation very attractive for WPT[158].

Finally, topologically protected (defected or edge) modes in electromagnetic analogues of topological insulators hold a grand promise for efficient and robust WPT[159–161]. These topologically nontrivial modes have been recently demonstrated for topologically protected unidirectional waveguides, topological LC-circuits, robust delay lines, and single-mode lasers[162]. In ref.[160], applying a Su-Schrieffer-Heeger (SSH) topological chain of magnetically coupled resonators for nonradiative WPT through an asymmetric topological edge state has been investigated. The topological protection of WPT in this system to random perturbations of the system's parameters was demonstrated. In another work[159], the advantage of PT-symmetry and topological edge states in photonic dimer chain composed of ultra-subwavelength resonators for long-range WPT system immune to the inner disorder perturbation was unveiled. Although topological systems allow reliable point-to-point energy transfer with immunity to disorder or geometric imperfections, they are inherently sensitive to absorption loss. Consequently, their expansion to the non-Hermitian regime is highly desirable. It can enable a novel form of stable topological waves[163] or make topological edge states robust to various non-Hermitian defects.

The unprecedented electromagnetic field control capabilities provided by metamaterials and metasurfaces have been compromised by the fact that these attributes are preset after the design and fabrication. Reconfigurable metasurfaces with versatile and dynamic functionality for WPT applications are highly desirable, but still largely unattainable. For this purpose, various types of non-linearity and tuning have to be investigated. A recently emerged concept of moiré metasurfaces, or stacking of two twisted metasurfaces[164], can become very fruitful to this end.

All these potential directions suggest that big opportunities to discover new physical concepts of WPT lie at the boundary between different approaches. We believe that further interpenetration of the concepts of nonradiative field control, time-reversal symmetry breaking, PT-symmetry breaking, nontrivial topology and new materials discussed in this work will inspire new breakthroughs in advanced WPT.

**Acknowledgements**

The work on the chapter New concepts of WPT was supported by the Russian Science Foundation (Project No. 20-72-10090). The work on the chapter Metamaterials and Metasurfaces for WPT was supported by the Russian Science Foundation (Project No. 21-79-30038). This work is partially supported by the Academy of Finland (Academy of Finland postdoctoral researcher grant 333479)

**Author contributions**

M.S., P.J., P.K., C.S., S.T., and A.K. conceived the project. A.K. managed the project. All authors contributed to the writing of the manuscript.

**Competing interests**

The authors declare no competing interests.